\documentclass[12pt]{article}

\usepackage{amsmath}

\def\eptwo{\left\{ \phantom{|}^{\mu\nu}_{ab} \right\}}
\def\epthree{\left\{ \phantom{|}^{\mu\nu\alpha}_{abc} \right\}}
\def\epfour{\left\{ \phantom{|}^{\mu\nu\alpha\beta}_{abcd} \right\}}
\def\epfive{\left\{ \phantom{|}^{\mu\nu\alpha\beta\gamma}_{abcde}
\right\}}

\title{Formalism of gauge invariant curvatures and constructing the
cubic vertices for massive spin-$\frac{3}{2}$ field in $AdS_4$ space}

\author{I.L. Buchbinder${}^{ab}$\thanks{joseph@tspu.edu.ru},
T.V. Snegirev${}^a$\thanks{snegirev@tspu.edu.ru}, Yu.M.
Zinoviev$^c$\thanks{Yurii.Zinoviev@ihep.ru}
\\[0.5cm]
\it ${}^a$Department of Theoretical Physics,\\
\it Tomsk State Pedagogical University,\\
\it Tomsk 634061, Russia\\[0.3cm]
\it ${}^b$National Research Tomsk State University, Russia\\[0.3cm]
\it ${}^c$Institute for High Energy Physics,\\
\it Protvino, Moscow Region, 142280, Russia}

\date{}

\begin{document}

\maketitle

\begin{abstract}
We study the interaction of massive spin-3/2 field with
electromagnetic and gravitational fields in the four dimensional
$AdS$ space and construct the corresponding cubic vertices.
Construction is based on generalization of Fradkin-Vasiliev
formalism, developed for massless higher spin fields, to massive
fermionic higher spin fields. The main ingredients of this formalism
are the gauge invariant curvatures. We build such curvatures for the
massive theory under consideration and show how the cubic vertices are
written in their terms.

\end{abstract}

\thispagestyle{empty}
\newpage
\setcounter{page}{1}

\newpage
\section*{Introduction}

Till now most of the investigations of consistent cubic interaction
vertices for higher spin fields were devoted to bosonic massless
fields (e.g.
\cite{Vas01,AV02,BFPT06,BL06,BLS08,Zin10,MMR10a,MMR10b,JT11} The
results on massive higher spin interactions are not so numerous,
although there exist classification of massive and massless cubic
vetices in flat space by Metsaev \cite{Met05,Met07b,Met12} as well
as some concrete examples
\cite{Zin06,Zin08,Zin09,ST10,BSZ12,Zin12,BSZ12b,BDT13,Zin13}. At the
same time there exists just a small number of papers devoted to
fermionic higher spin interactions \cite{Met06,Met07b,GHR12,HGR13}.
At least one of the reason for this is that technically such
investigations appears to be much more involved. In this paper,
using massive spin $\frac{3}{2}$ as a simple but physically
interesting and non-trivial example of massive fermionic higher spin
fields, we apply the so-called Fradkin-Vasiliev formalism
\cite{FV87,FV87a} (see also
\cite{Alk10,Zin10a,BSZ11,Zin11,BS11,BPS12,Zin13})  to the
construction of electromagnetic and gravitational cubic vertices.

Let us briefly remind basic properties of this formalism. Higher spin
particle is describes by a set of fields that we collectively denote
as $\Phi$ here and for each field one can construct gauge invariant
object that we will call curvature and denote as ${\cal R}$. Moreover
with help of these curvatures free Lagrangian can be rewritten in
explicitly gauge invariant form as:
$$
{\cal L}_0 \sim \sum {\cal R} \wedge {\cal R}
$$
Using these ingredients one can construct two types of
non-trivial\footnote{We call vertex trivial if it can be constructed
using gauge invariant curvatures only, i.e. ${\cal L} \sim {\cal R}
\wedge {\cal R} \wedge {\cal R}$} cubic vertices:
\begin{itemize}
\item Abelian vertices that have the form:
$$
{\cal L} \sim {\cal R} \wedge {\cal R} \wedge \Phi
$$
\item non-Abelian ones that look like:
$$
{\cal L} \sim {\cal R} \wedge \Phi \wedge \Phi
$$
\end{itemize}
For the massless bosonic fields Vasiliev \cite{Vas11} has shown that
any such non-Abelian vertex can be obtained as a result of
deformation procedure that can be described  as follows.
\begin{itemize}
\item Construct quadratic in fields deformations for the curvatures
and linear corrections to the gauge transformations
\begin{equation}\label{Step1}
\Delta {\cal R} \sim \Phi \wedge \Phi, \qquad \delta_1 \Phi \sim
\Phi \xi
\end{equation}
so that deformed curvatures
$\hat{\cal{R}}={\cal{R}}+\Delta{\cal{R}}$ transform covariantly
\begin{equation}\label{Step2}
\delta \hat{\cal R} = \delta_1 {\cal R} + \delta_0 \Delta {\cal R}
\sim {\cal R} \xi
\end{equation}
At this step most of the arbitrary parameters are fixed, however,
there still remains some ambiguity. The reason is that covariance of
transformations for deformed curvatures guarantees that the
equations of motion will be gauge invariant but it does not
guarantee that they will be Lagrangian.

\item Non-Abelian cubic vertex arises then one put these deformed
curvatures into the free Lagrangian and using explicit form of the
gauge transformations requires it to be gauge invariant
\begin{eqnarray*}
{\cal{L}}\sim\sum\hat{\cal{R}}\hat{\cal{R}}\quad\Leftrightarrow\quad
\delta{\cal{L}}\sim\sum{\cal{R}}{\cal{R}}\xi=0
\end{eqnarray*}
\end{itemize}
Note at last that Vasiliev has shown \cite{Vas11} that any
non-trivial cubic vertex for three fields with spins $s_1$, $s_2$
and $s_3$ having up to $s_1+s_2+s_3-2$ derivatives can be
constructed as some combination of Abelian and non-Abelian vertices.
However, ones point out that the Vasiliev construction \cite{Vas11}
was initially formulated only for massless higher spin field
theories. In this paper we generalize the Vasiliev approach to
massive higher spin theories on the example of cubic electromagnetic
and gravitational coupling for massive spin-$\frac{3}{2}$ field in
the $AdS_4$ space.

The paper is organized as follows. In  Section 1 we give all necessary
kinematics formulae. Here we introduce two gauge invariant objects
(for the physical field $\psi_\mu$ and Stueckelberg one $\phi$) and
rewrite the free Lagrangian using these objects. Section 2 devoted to
the electromagnetic interaction for such massive spin-$\frac{3}{2}$
field (for previous results see e.g. \cite{DPW00,DW01d,Rah11}).
 In subsection 2.1, as a independent check for our calculations
as well as for instructive comparison, we consider this task using
straightforward constructive approach. Then in subsection 2.2 we
consider the same task in the Fradkin-Vasiliev formalism. At last in
Section 3 we construct gravitational cubic vertex in the
Fradkin-Vasiliev formalism.

\noindent
{\bf Notations and conventions}

We work in four-dimensional Anti-de Sitter space with cosmological
constant $\Lambda$. The $AdS$ covariant derivatives $D_\mu$ is
normalized so that\footnote{We use Greek letters $\mu,\nu,...$ for
world indices and Latin letters $a,b,...$ for local ones. Summation
over any repeated indices is implied. For indices in (square) round
brackets we use convention of complete (anti)-symmetrization without
normalization factor. Spinor indices of (tensor)-spinor fields are
omitted.}
\begin{equation}\label{AdSDerivative}
[D_{\mu}, D_{\nu}] \xi^a = \lambda^2 (e_{[\mu}{}^a \xi_{\nu]} +
\frac12 \Gamma_{\mu\nu} \xi^a), \qquad \lambda^2 = - \frac{\Lambda}{3}
\end{equation}

In (\ref{AdSDerivative}) $e_\mu{}^a$ is background (non-dynamical)
tetrad linking world and local indexes and has a standard definition
$e_\mu{}^ae_\nu{}^bg^{ab}=g_{\mu\nu}$, where $g_{\mu\nu}$ and
$g^{ab}$ are respectively curved world $AdS$ metrics and flat local
one. Wherever it is convenient local indices are converted into the
world ones by $e_\mu{}^a$ and its inverse $e^\mu{}_a$, in particular
the matrix $\gamma_\mu$ with world index is understood as
$\gamma_\mu = e_\mu{}^a \gamma^a$. Also we introduce the following
notation for antisymmetric combinations of $e_\mu{}^a$
$$
\eptwo = e^{[\mu}{}_a e^{\nu]}{}_b, \qquad
\epthree = e^{[\mu}{}_a e^\nu{}_b e^{\alpha]}{}_c, \qquad
\epfour = e^{[\mu}{}_a e^\nu{}_b e^\alpha{}_c e^{\beta]}{}_d
$$

In expression (\ref{AdSDerivative})
$\Gamma^{ab}=\frac12\gamma^{[a}\gamma^{b]}$ is antisymmetric
combination of two gamma matrices and is a particular case of the more
general definition which we will use
$$
\Gamma^{a_1a_2...a_n} = \frac{1}{n!} \gamma^{[a_1} \dots \gamma^{a_n]}
$$
where $n=2,3,4$ for four-dimensional space. Let us present
the main properties of such $\Gamma$-matrices
\begin{eqnarray}\label{Properties}
\Gamma^{ab_1...b_n}&=&-g^{a[b_1}\Gamma^{b_2...b_n]}+\gamma^a\Gamma^{b_1...b_n}
\nonumber\\
\Gamma^{ab_1...b_n}&=&g^{a[b_1}\Gamma^{b_2...b_n]}+(-1)^n
\Gamma^{b_1...b_n}\gamma^a
\nonumber\\
\gamma^a\Gamma^{b_1...b_n}&=&2g^{a[b_1}\Gamma^{b_2...b_n]}
+(-1)^n\Gamma^{b_1...b_n}\gamma^a
\\
 \gamma_a\Gamma^{ab_1...b_n}&=&(d-n)\Gamma^{b_1...b_n}
\nonumber\\
 \Gamma^{ab_1...b_n}\gamma_a&=&(-1)^n(d-n)\Gamma^{b_1...b_n}\nonumber
\end{eqnarray}
The identities (\ref{Properties}) are used in the paper to verify
the gauge invariance of the Lagrangians.

All fermionic fields are anticommuting and Majorana. We also use
Majorana representation of $\gamma$-matrices with hermitian
conjugation defined as follows:
\begin{eqnarray*}
(\gamma^a)^\dag &=& \gamma^0 \gamma^a \gamma^0 \\
(\Gamma^{a_1a_2...a_n})^\dag &=& \gamma^0 \Gamma^{a_n...a_2a_1}
\gamma^0 = - \gamma^0 \Gamma^{a_1a_2...a_n} \gamma^0
\end{eqnarray*}
In this case $\gamma^0\gamma^a$ and $\gamma^0\Gamma^{ab}$ are
symmetric in spinor indices while $\gamma^0$, $\gamma^0\Gamma^{abc}$,
and $\gamma^0\Gamma^{abcd}$ are antisymmetric.

\section{Kinematic of massive spin-3/2 field}

For gauge-invariant description of massive spin-3/2 field besides
the master vector-spinor field $\Psi_\mu$ one has also introduce
Stueckelberg spinor field $\phi$. The free Lagrangian is known and
has the form:
\begin{eqnarray}\label{FreeLagMassvS3/2}
{\cal L}_0 &=& -\frac{i}{2} \epthree \bar{\psi}_\mu \Gamma^{abc} D_\nu
\psi_\alpha + \frac{i}{2} e^\mu{}_a \bar{\phi} \gamma^a D_\mu \phi
\nonumber\\
&& + 3im e^\mu{}_a \bar{\psi}_\mu \gamma^a \phi - \frac{3M}{2} \eptwo
\bar{\psi}_\mu \Gamma^{ab} \Psi_\nu - M \bar{\phi} \phi
\end{eqnarray}
This Lagrangian is invariant under the following gauge transformations
\begin{eqnarray*}
\delta_0 \psi_\mu &=& D_\mu \xi + \frac{iM}{2} \gamma_\mu \xi \\
\delta_0 \phi &=& 3m \xi
\end{eqnarray*}
where $M^2=m^2+\lambda^2 = m^2 - \Lambda$ and $m$ is a mass parameter.
Note that such description works in de Sitter space as well provided
$m^2 > \Lambda$, $m^2 = \Lambda$ being the boundary of unitarity
region.

Using the explicit form of gauge transformations one can construct two
gauge-invariant objects (curvatures):
\begin{eqnarray}\label{CurvMassvS3/2}
\Psi_{\mu\nu} &=& D_{[\mu} \psi_{\nu]} + \dfrac{m}{6} \Gamma_{\mu\nu}
\phi + \dfrac{iM}{2} \gamma_{[\mu} \psi_{\nu]} \nonumber\\
\Phi_\mu &=& D_\mu \phi - 3m \psi_\mu + \dfrac{iM}{2} \gamma_\mu \phi
\end{eqnarray}
which satisfy the Bianchi identities
\begin{eqnarray}\label{BianIdentMassvS3/2}
D_{[\mu} \Psi_{\nu\alpha]} &=& \dfrac{m}{6} \Gamma_{[\mu\nu}
\Phi_{\alpha]} - \dfrac{iM}{2} \gamma_{[\mu} \Psi_{\nu\alpha]}
\nonumber\\
D_{[\mu} \Phi_{\nu]} &=& - 3m \Psi_{\mu\nu} - \dfrac{iM}{2}
\gamma_{[\mu} \Phi_{\nu]}
\end{eqnarray}
Note that curvatures (\ref{CurvMassvS3/2}) are closely related with
the Lagrangean equations of motion. It can be seen from the total
variation of Lagrangian (\ref{FreeLagMassvS3/2}) that can be written
as follows:
\begin{eqnarray}\label{VarMassvS3/2}
\delta {\cal L }_0 &=& -\frac{i}{2} \epthree \bar{\Psi}_{\mu\nu}
\Gamma^{abc} \delta \psi_\alpha - i e^\mu{}_a \bar{\Phi}_\mu \gamma^a
\delta \phi
\end{eqnarray}

Moreover using these curvatures the free Lagrangian
(\ref{FreeLagMassvS3/2}) can be rewritten in explicitly
gauge-invariant form. The most general ansatz looks like:
\begin{equation}\label{CurvLagMassvS3/2}
{\cal L}_0 = c_1 \epfour \bar{\Psi}_{\mu\nu} \Gamma^{abcd}
\Psi_{\nu\alpha} + ic_2 \epthree \bar{\Psi}_{\mu\nu}
\Gamma^{abc}\Phi_\alpha + c_3 \eptwo \bar{\Phi}_\mu \Gamma^{ab}
\Phi_\nu
\end{equation}
The requirement to reproduce the original Lagrangian
(\ref{FreeLagMassvS3/2}) partially fixes the parameters
\begin{equation}\label{CondLagParam}
3c_3 = - 8c_1, \qquad 6c_2m = \frac{1}{2} - 16c_1M
\end{equation}
The remaining freedom in parameters is related with the identity:
$$
\epfour D_{\mu} [\bar{\Psi}_{\nu\alpha} \Gamma^{abcd} \Phi_\beta] =
0
$$
Using the Bianchi identities for the curvatures
(\ref{BianIdentMassvS3/2}) we obtain
$$
-  3m \epfour \bar{\Psi}_{\mu\nu} \Gamma^{abcd} \Psi_{\nu\alpha}
+ 8iM \epthree \bar{\Psi}_{\mu\nu} \Gamma^{abc} \Phi_\alpha
+ 8m \eptwo \bar\Phi_\mu \Gamma^{ab} \Phi_\nu = 0
$$
As it will be seen in the next section sometimes this ambiguity may be
important in order to reproduce the most general cubic vertex so we
will not fix it here.

\section{Electromagnetic interaction}

\subsection{Constructive approach}

We prefer to work with Majorana fermions thus for the description of
electromagnetic interactions we will use pair of them
$\psi_\mu{}^i$, $\phi^i$, where $i=1,2$ is the $SO(2)$ index. Let us
switch minimal electromagnetic interaction by standard rule
$$
D_\mu \psi_\nu{}^i \Rightarrow D_\mu \psi_\nu{}^i + e_0
\varepsilon^{ij} A_\mu \psi_\nu{}^j
$$
and similarly for $\phi^i$. The Lagrangian (\ref{FreeLagMassvS3/2})
with above replacement for covariant derivative is not invariant
under the free gauge transformations, variation of the Lagrangian
has the form:
\begin{equation}
\delta_0 {\cal L}_0 = - \frac{ie_0}{2} \varepsilon^{ij} \epthree
\bar{\psi}_\mu{}^i \Gamma^{abc} F_{\nu\alpha} \xi^j = - 3ie_0
\varepsilon^{ij} e^\mu{}_a \bar{\psi}_\mu{}^i \Gamma^{abc} F^{bc}
\xi^j
\end{equation}
To compensate for this non-invariance let us introduce non-minimal
interactions as follows:
\begin{eqnarray}\label{emvertex}
{\cal L}_1&=&\varepsilon^{ij} [ \eptwo \bar{\psi}_\mu{}^i ( b_1
F^{ab} + b_2 \Gamma^{abcd} F^{cd}) \psi_\nu{}^j  \nonumber\\
&&+i e^\mu{}_a \bar{\psi}_\mu{}^i (b_3 F^{ab} \gamma^b + b_4
\Gamma^{abc} F^{bc}) \phi^j + b_5 \bar{\phi}^i (\Gamma F) \phi^j ]
\end{eqnarray}
as well as corresponding corrections to gauge transformations:
\begin{eqnarray}
\delta_1 \psi_\mu{}^i &=& i\alpha_1 \varepsilon^{ij} (\Gamma F)
\gamma_\mu \xi^i, \qquad \delta \phi^i = \alpha_2 \varepsilon^{ij}
(\Gamma F) \xi^j \nonumber \\
\delta_1 A_\mu &=& \beta_1 \varepsilon^{ij} (\bar{\psi}_\mu{}^i
\xi^j) + i\beta_2 \varepsilon^{ij} (\bar{\phi}^i \gamma_\mu \xi^j)
\end{eqnarray}
By direct calculations one can check that gauge invariance can be
restored that gives the solution for arbitrary coefficients:
$$
b_1 = 6\alpha_1, \quad b_2 = 3\alpha_1, \quad b_3 = - 2\alpha_2,
\quad b_4 = \alpha_2, \quad b_5 = \frac{M\alpha_2}{3m},\quad \beta_1
= - 24\alpha_1, \quad \beta_2 = - 2\alpha_2
$$
provided the following important relation holds:
\begin{equation}
4M\alpha_1 + 2m\alpha_2 = e_0
\end{equation}
Lagrangian (\ref{emvertex}) together with correction, stipulated by
minimal switch of interaction in free Lagrangian, is the final cubic
electromagnetic interaction vertex for spin-3/2 field. As we see,
this vertex contains two arbitrary parameters.

A few comments are in order.
\begin{itemize}
\item Calculating commutator of two gauge transformations we obtain
$$
[ \delta_1, \delta_2 ] A_\mu = - 8\varepsilon^{ij} (6\alpha_1{}^2 +
\alpha_2{}^2) (\xi_2{}^i \gamma^\nu \xi_1{}^j) F_{\mu\nu}
$$
and it means that for non-zero electric charge $e_0$ any such model
must be component of some (spontaneously broken) supergravity.
\item From the supergravity point of view the remaining freedom in the
parameters $\alpha_1$ and $\alpha_2$ is clear: in general our vector
field is a superposition of graviphoton (i.e. vector field from the
supermultiplet $(\frac{3}{2},1)$) and goldstino's superpartner (i.e.
vector field from the supermultiplet $(1,\frac{1}{2})$.
\item From the expression for the parameter $b_5$ one can see that we
have an ambiguity between flat and massless limits. Indeed, in the
flat case $b_5 = \frac{\alpha_2}{3}$ and does not depend on the mass
any more, while for the non-zero cosmological term this parameter is
singular in the massless limit.
\item The most simple case --- $\alpha_2 = 0$, i.e. vector field is
just graviphoton. In this case electric charge $e_0 = \frac{2}{3}
M\alpha_1$ becomes zero at the boundary of the unitarity region.
\end{itemize}

\subsection{Formulation in terms of curvatures}

Now we reformulate the results of the subsection {\bf 2.1} on the
base of Fradkin-Vasiliev formalism. Following general procedure we
begin with deformation of curvatures so that they transform
covariantly (\ref{Step1}), (\ref{Step2}). In the case under
consideration
${\cal{R}}=\{\Psi_{\mu\nu}{}^i,\Phi_\mu{}^i,F_{\mu\nu}\}$. For the
fermionic curvatures deformations correspond simply to the minimal
substitution:
\begin{eqnarray}
\Delta \Psi_{\mu\nu}{}^i &=& e_0 \varepsilon^{ij} A_{[\mu}
\psi_{\nu]}{}^j \nonumber \\
\Delta \Phi_\mu{}^i &=& e_0 \varepsilon^{ij} A_\mu \phi^j
\end{eqnarray}
while transformations for the deformed curvatures look like:
\begin{equation}\label{EmCurvTransfMassvS3/2}
\delta \hat{\Psi}_{\mu\nu}{}^i = e_0 \varepsilon^{ij} F_{\mu\nu}
\xi^j, \qquad \delta \hat{\Phi}_\mu{}^i = 0
\end{equation}
where
$\hat{\Psi}_{\mu\nu}{}^i = \Psi_{\mu\nu}{}^i +
\Delta \Psi_{\mu\nu}{}^i$, $\hat{\Phi}_\mu{}^i = \Phi_\mu{}^i +
\Delta \Phi_\mu{}^i$ and $\Psi_{\mu\nu}{}^i$, $\Phi_\mu{}^i$ are given
by (\ref{CurvMassvS3/2}).

The most general deformation for the electromagnetic fields strength
quadratic in fermionic fields can be written as follows:
\begin{equation}
\Delta F_{\mu\nu} = \varepsilon^{ij} [a_1 \bar\psi_{[\mu}{}^i
\psi_{\nu]}{}^j + ia_2 \bar\psi_{[\mu}{}^i \gamma_{\nu]} \phi{}^j +
a_3 \phi^i \Gamma_{\mu\nu} \phi^j]
\end{equation}
Corresponding corrections to the gauge transformation have the form:
\begin{equation}
\delta_1 A_\mu = \varepsilon^{ij} [2a_1 \bar{\psi}_\mu{}^i \xi^j -ia_2
\bar{\phi}^i \gamma_\mu \xi^j]
\end{equation}
The deformed curvatures will transform covariantly
\begin{equation}\label{MassvCurvTransfS1}
\delta_1 \hat{F}_{\mu\nu} = \varepsilon^{ij} [2a_1
\bar{\Psi}_{\mu\nu}^i \xi^j - ia_2 \bar{\Phi}_{[\mu}{}^i
\gamma_{\nu]} \xi^j]
\end{equation}
provided the following relation holds:
\begin{equation}
3ma_3 = - \frac{ma_1}{6} - Ma_2
\end{equation}
Here $\hat{F}_{\mu\nu}=F_{\mu\nu}+\Delta F_{\mu\nu}.$

Now let us consider the Lagrangian
\begin{eqnarray}
{\cal L}_1 &=& - \frac{1}{4} \hat{F}^{\mu\nu} \hat{F}_{\mu\nu} +
c_1 \epfour \hat{\bar{\Psi}}_{\mu\nu}{}^i \Gamma^{abcd}
\hat{\Psi}_{\alpha\beta}{}^i \nonumber \\
&& + ic_2 \epthree \hat{\bar{\Psi}}_{\mu\nu}{}^i \Gamma^{abc}
\hat{\Phi}_\alpha{}^i + c_3 \eptwo \hat{\bar\Phi}_\mu{}^i \Gamma^{ab}
\hat\Phi_\nu{}^i
\end{eqnarray}
where all initial curvatures (including electromagnetic field
strength) are replaced by the deformed ones, and require that this
Lagrangian be invariant. Non-trivial variations arise only with
respect to transformations with the spinor parameter $\xi^i$. Using
explicit form of the covariant transformations
(\ref{EmCurvTransfMassvS3/2}),
(\ref{MassvCurvTransfS1}) we obtain
\begin{eqnarray}\label{EmVarMassvS3/2}
\delta{\cal L} &=& - \frac{1}{2} \varepsilon^{ij} \eptwo
\bar{\Psi}_{\mu\nu}{}^i (a_1 F^{ab} - 48e_0c_1 \Gamma^{abcd} F^{cd})
\xi^j  \nonumber\\
&& + i \varepsilon^{ij} e^\mu{}_a \bar{\Phi}_{\mu}{}^i (a_2 \gamma^{b}
F^{ab} + 6e_0c_2 \Gamma^{abc} F^{bc}) \xi^j
\end{eqnarray}
To compensate these terms one need to introduce non-minimal
corrections (note that they have exactly the same form as in the
previous subsection):
\begin{equation}
\delta_1 \psi_\mu{}^i = i \alpha_1 \varepsilon^{ij} (\Gamma F)
\gamma_\mu \xi^j, \qquad \delta_1 \phi^i = \alpha_2
\varepsilon^{ij}(\Gamma F) \xi^j
\end{equation}
which in turn produce the following variations:
\begin{eqnarray}
\delta_1 {\cal L}_0 &=& - \alpha_1 \varepsilon^{ij} \eptwo
\bar{\Psi}_{\mu\nu}{}^i (6 F^{ab} + 3 F^{de} \Gamma^{abde}) \xi^j
\nonumber \\
&& - i\alpha_2 \varepsilon^{ij} e^\mu{}_a \bar{\Phi}_\mu{}^i (2
\gamma^b F^{ab} + \Gamma^{abc} F^{bc}) \xi^j
\end{eqnarray}
Comparing with (\ref{EmVarMassvS3/2}) we conclude
\begin{equation}\label{CondEmIntParam}
a_1 = - 12\alpha_1, \qquad a_2 = 2\alpha_2, \qquad
\alpha_1 = 8e_0c_1,\qquad \alpha_2 = 6e_0c_2
\end{equation}
Thus we see that the choice of the parameters $\alpha_1$ and
$\alpha_2$ is related with the choice of parameters $c_1$ and $c_2$ in
the free Lagrangian. Moreover, if we use the relation
$$
6mc_2 = \frac{1}{2} - 16Mc_1
$$
we again obtain
$$
4M\alpha_1 + 2m\alpha_2 = e_0
$$

Let us extract the cubic vertex. Using relations
(\ref{CondLagParam}), (\ref{CondEmIntParam}) we get
\begin{eqnarray}\label{emavertex}
{\cal L}_1 &=& 3\alpha_1 \varepsilon^{ij} \eptwo \bar{\psi}_{\mu}{}^i
(2\alpha_1 F^{ab} + \Gamma^{abcd} F^{cd}) \psi_{\nu}{}^j  \nonumber
\\
&& - i\alpha_2 \varepsilon^{ij} e^\mu{}_a \bar{\psi}_{\mu}{}^i
(2\gamma^b F^{ab} - \Gamma^{abc} F^{bc}) \phi^j + \frac{M\alpha_2}{3m}
\varepsilon^{ij} \phi^i (\Gamma F) \phi^j  \nonumber \\
&& + \frac{ie_0}{2} \varepsilon^{ij} \epthree A_{\mu}
\bar{\psi}_{\nu}{}^i \Gamma^{abc} \psi_{\alpha}{}^j +\frac{ie_0}{2}
\varepsilon^{ij} e^\mu{}_a A_\mu \bar{\phi}^i \gamma^{a} \phi^j
\end{eqnarray}
Here the last line corresponds to the minimal interactions while the
other terms are non-minimal corrections.

The Lagrangian ${\cal L}_1$ (\ref{emavertex}) up to the minimal
interaction is the same one obtained in the subsection ${\bf 2.1}$
for the electromagnetic cubic vertex.

\section{Gravitational interaction}

\subsection{Kinematics for gravity}

Let us briefly review basic features of gravity in the $AdS_4$ space
at free level. In the frame formulation the gravitational field is
described by dynamical frame $h_\mu{}^a$ and Lorentz connection
$\omega_\mu{}^{ab}$ being antisymmetric in local indices. Free
Lagrangian in $AdS$ space have the form
\begin{equation}\label{FreeLagS2}
{\cal L}_0 = \frac{1}{2} \eptwo \omega_\mu{}^{ac} \omega_{\nu}{}^{bc}
- \frac{1}{2} \epthree \omega_\mu{}^{ab} D_\nu h_\alpha{}^c +
\lambda^2 \eptwo h_\mu{}^a h_\nu{}^b
\end{equation}
and is invariant under the gauge transformations
\begin{eqnarray}
\delta_0 \omega_\mu{}^{ab} &=& D_\mu \hat{\eta}^{ab} - \lambda^2
e_\mu{}^{[a} \hat{\xi}^{b]} \nonumber \\
\delta_0 h_\mu{}^a &=& D_\mu \hat{\xi}^a + \hat{\eta}_\mu{}^{a}
\end{eqnarray}
The gauge invariant objects (linearized curvature and torsion)
have the form
\begin{eqnarray}\label{CurvTors}
R_{\mu\nu}{}^{ab} &=& D_{[\mu} \omega_{\nu]}{}^{ab} - \lambda^2
e_{[\mu}{}^{[a} h_{\nu]}{}^{b]} \nonumber\\
T_{\mu\nu}{}^a &=& D_{[\mu} h_{\nu]}{}^a - \omega_{[\mu,\nu]}{}^{a}
\end{eqnarray}
They satisfy the Bianchi identities
\begin{eqnarray}\label{BianIdentS2}
D_{[\mu} R_{\nu\alpha]}{}^{ab} &=& \lambda^2 e_{[\mu}{}^{[a}
T_{\nu\alpha]}{}^{b]} \nonumber\\
D_{[\mu} T_{\nu\alpha]}{}^a &=& - R_{[\mu\nu,\alpha]}{}^{a}
\end{eqnarray}
Note that on the mass shell for the auxiliary field
$\omega_\mu{}^{ab}$ by virtue of (\ref{BianIdentS2}) we have
\begin{equation}\label{OnShellS2}
T_{\mu\nu}{}^a = 0 \quad\Rightarrow\quad R_{[\mu\nu,\alpha]}{}^{a} =
0,\qquad D_{[\mu} R_{\nu\alpha]}{}^{ab} = 0
\end{equation}
At last the free Lagrangian can be rewritten as follows
\begin{equation}\label{CurvLagS2}
{\cal L}_0 = c_0 \epfour R_{\mu\nu}{}^{ab} R_{\alpha\beta}{}^{cd},
\qquad c_0 = \frac{1}{32\lambda^2}
\end{equation}

\subsection{Gravitational coupling for massive spin-3/2}

Following general scheme (\ref{Step1}), (\ref{Step2}) we begin with
the deformation of the curvatures that in the case under
consideration are
${\cal{R}}=\{\Psi_{\mu\nu},\Phi_\mu,R_{\mu\nu}{}^{ab},
T_{\mu\nu}{}^a\}$.
As in the case of electromagnetic interactions deformations for
spin-3/2 curvatures correspond to the minimal substitution rules,
i.e. to the replacement of covariant derivative $D\rightarrow
D+\omega$ and background tetrad $e_\mu{}^a\rightarrow e_\mu{}^a+
h_\mu{}^a$:
\begin{eqnarray}
\Delta {\Psi}_{\mu\nu} &=& a_1 (\omega_{[\mu}{}^{ab} \Gamma_{ab}
\psi_{\nu]} + 2Mi h_{[\mu}{}^a \gamma_a \psi_{\nu]} -
\frac{2m}{3} h_{[\mu}{}^a \Gamma_{\nu]}{}^{a} \phi) \nonumber \\
\Delta \Phi_\mu &=& a_1 (\omega_\mu{}^{ab} \Gamma_{ab} \phi + 2Mi
h_\mu{}^a \gamma_a \phi)
\end{eqnarray}
Corrections to the gauge transformations will look like:
\begin{eqnarray}
\delta_1 \Psi_\mu &=& - a_1 (\Gamma^{ab} \Psi_\mu \hat{\eta}_{ab}
+ 2iM \gamma^a \psi_\mu \hat{\xi}_a - \frac{2m}{3} \Gamma_\mu{}^{a}
\phi \hat{\xi}_a \nonumber\\
&& - \omega_\mu{}^{ab} \Gamma_{ab} \xi - 2iM h_\mu{}^a \gamma_a \xi)
\\
\delta_1 \phi &=& - a_1 (\Gamma^{ab} \phi \hat{\eta}_{ab} + 2iM
\gamma^a \phi \hat{\xi}_a) \nonumber
\end{eqnarray}
while transformations for the deformed curvatures will have the form:
\begin{eqnarray}\label{GrCurvTransfMassvS3/2}
\delta \hat{\Psi}_{\mu\nu} &=& -a_1 (\Gamma^{ab} \Psi_{\mu\nu}
\hat{\eta}_{ab} + 2iM \gamma^a \Psi_{\mu\nu} \hat{\xi}_a +
\frac{2m}{3} \Gamma_{[\mu}{}^{a} \Phi_{\nu]} \hat{\xi}_a \nonumber\\
&& - R_{\mu\nu}{}^{ab} \Gamma_{ab} \xi - 2iM T_{\mu\nu}{}^a \gamma_a
\xi) \\
\delta \hat{\Phi}_\mu &=& -a_1 (\Gamma^{ab} \Phi_\mu \eta_{ab} + 2iM
\gamma^a \Phi_\mu \hat{\xi}_a) \nonumber
\end{eqnarray}
Here $\hat{\Psi}_{\mu\nu}={\Psi}_{\mu\nu}+\Delta {\Psi}_{\mu\nu},
\hat{\Phi}_\mu={\Phi}_\mu+\Delta \Phi_\mu$ and ${\Psi}_{\mu\nu}{}^i,
{\Phi}^i$ are given by (\ref{CurvMassvS3/2}).

Now let us consider the most general deformations for gravitational
curvature and torsion quadratic in spin-3/2 fields:
\begin{eqnarray}
\Delta R_{\mu\nu}{}^{ab} &=& b_1 \bar{\psi}_{[\mu} \Gamma^{ab}
\psi_{\nu]} + ib_2 e_{[\mu}{}^{[a} \bar{\psi}_{\nu]} \gamma^{b]} \phi
+ ib_3 \bar{\psi}_{[\mu} \Gamma_{\nu]}{}^{ab} \phi + b_4 e_{[\mu}{}^a
e_{\nu]}{}^b \bar{\phi} \phi + b_5 \bar{\phi} \Gamma_{\mu\nu}{}^{ab}
\phi \nonumber \\
\Delta T_{\mu\nu}{}^a &=& ib_6 \bar{\psi}_{[\mu} \gamma^{a}
\psi_{\nu]} + b_7 e_{[\mu}{}^a \bar{\psi}_{\nu]} \phi + b_8
\bar{\psi}_{[\mu} \Gamma_{\nu]}{}^{a} \phi + ib_9 \bar{\phi}
\Gamma_{\mu\nu}{}^{a} \phi
\end{eqnarray}
Note that there are three possible field redefinitions
\begin{equation}\label{FieldRedef}
\omega_\mu{}^{ab} \Rightarrow \omega_\mu{}^{ab} + i\kappa_1 \bar{\phi}
\Gamma_\mu{}^{ab} \phi, \qquad
h_\mu{}^a \Rightarrow h_\mu{}^a + \kappa_2 \bar{\psi}_\mu \gamma^a
\phi + \kappa_3 e_\mu{}^a \bar{\phi} \phi
\end{equation}
that shift parameters $b_3,b_6$ and $b_7$.

In order that deformed curvatures transform covariantly we have to
introduce the following corrections to the gauge transformations:
\begin{eqnarray}
\delta_1 \omega_\mu{}^{ab} &=& 2b_1 \bar{\psi}_\mu \Gamma^{ab} \xi
-ib_2 e_\mu{}^{[a} \bar{\phi} \gamma^{b]} \xi - ib_3 \bar{\phi}
\Gamma_\mu{}^{ab} \xi \nonumber \\
\delta_1 h_\mu{}^a &=& 2ib_6 \bar{\psi}_\mu \gamma^a \xi + b_7
e_\mu{}^a \bar{\phi} \xi + b_8 \bar{\phi} \Gamma_\mu{}^a \xi
\end{eqnarray}
Then the curvature and torsion will transform as follows:
\begin{eqnarray}
\delta \hat{R}_{\mu\nu}{}^{ab} &=& 2b_1 \bar{\Psi}_{\mu\nu}
\Gamma^{ab} \xi + ib_2 e_{[\mu}{}^{[a} \bar{\Phi}_{\nu]} \gamma^{b]}
\xi - ib_3 \bar{\Phi}_{[\mu} \Gamma_{\nu]}{}^{ab} \xi \nonumber \\
\delta \hat{T}_{\mu\nu}{}^a &=& 2ib_6 \bar{\Psi}_{\mu\nu} \gamma^a \xi
- b_7 e_{[\mu}{}^a \bar{\Phi}_{\nu]} \xi + b_8 \bar{\Phi}_{[\mu}
\Gamma_{\nu]}{}^a \xi
\end{eqnarray}
provided the following restrictions on the arbitrary parameters
hold:
\begin{align}\label{CondGrIntParam}
& 3mb_2=b_6(M^2-m^2)-b_1M && 3mb_8=-b_6M+b_1
\nonumber\\
& 6mb_4=\frac{mb_1}{3}+2b_7(M^2-m^2) && 6mb_9=-\frac{mb_6}{3}-2b_3
\\
& 6mb_5=-\frac{mb_1}{3}-2b_3M\nonumber
\end{align}
Here $\hat{R}_{\mu\nu}{}^{ab}={R}_{\mu\nu}{}^{ab}+\Delta
R_{\mu\nu}{}^{ab}, \hat{T}_{\mu\nu}{}^a={T}_{\mu\nu}{}^a+\Delta
T_{\mu\nu}{}^a$ and ${R}_{\mu\nu}{}^{ab}, {T}_{\mu\nu}{}^a$ are
given by (\ref{CurvTors}).

 General solution to these relations has four free parameters,
for-example, $b_{1,3,6,7}$. But as we have already noted three of
them are related with possible field redefinitions, so we have one
non-trivial parameter $b_1$ only.

Let us consider the following Lagrangian:
\begin{eqnarray}\label{gvertex}
{\cal L}&=&c_1 \epfour \hat{\bar{\Psi}}_{\mu\nu} \Gamma^{abcd}
\hat{\Psi}_{\nu\alpha} + ic_2 \epthree \hat{\bar{\Psi}}_{\mu\nu}
\Gamma^{abc} \hat{\Phi}_\alpha + c_3 \eptwo \hat{\bar{\Phi}}_\mu
\Gamma^{ab} \hat{\Phi}_\nu  \\
&& + c_0 \epfour \hat{R}_{\mu\nu}{}^{ab}
\hat{R}_{\alpha\beta}{}^{cd} + ic_4 \epfour \bar{\Psi}_{\mu\nu}
\Gamma^{abc} \Phi_{\alpha} h_\beta{}^d + c_5 \epthree \bar{\Phi}_\mu
\Gamma^{ab} \Phi_\nu h_\alpha{}^c\nonumber
\end{eqnarray}
where the first four terms are just the sum of free Lagrangians for
massive spin-$\frac{3}{2}$ and massless spin-2 with the initial
curvatures replaced by the deformed ones while the last two terms
are possible Abelian vertices. Note that in dimensions $d > 4$ we
would have to introduce one more Abelian vertex:
$$
\epfive \bar{\Psi}_{\mu\nu} \Gamma^{abcd} \Psi_{\alpha\beta}
h_\gamma{}^e
$$
Now we require that this Lagrangian be gauge invariant. To fix all
coefficients it is enough to consider variations that do not vanish
on-shell. For $\xi$-transformations we have
\begin{eqnarray*}
\delta{\cal L} &=& (-24c_1a_1+4b_1c_0) \epfour
\bar{\Psi}_{\alpha\beta} R_{\mu\nu}{}^{ab} \Gamma^{cd} \xi  \\
&& + i(6c_2a_1-8c_0b_3) \epthree R_{\mu\nu}{}^{ad} \bar{\xi}
\Gamma^{bcd} \Phi_\alpha + i(6c_2a_1-8c_0b_2) \epthree
R_{\mu\nu}{}^{ab} \bar{\xi} \gamma^c \Phi_\alpha
\end{eqnarray*}
the last two terms vanish on-shell and for the second term it can be
seen from the identity:
$$
0 = \epfour R_{\mu\nu,\alpha}{}^{a} \bar{\xi} \Gamma^{bcd} \Phi_\beta
= 3 \epthree R_{\mu\nu}{}^{ad} \bar{\xi} \Gamma^{bcd} \Phi_\alpha
$$
The first term does not vanish on-shell so we have to put:
\begin{equation}\label{CondGrIntParamMassv1}
b_1c_0 = 6c_1a_1
\end{equation}
Calculating variations for $\hat{\eta}^{ab}$-transformations we
obtain:
\begin{eqnarray*}
\delta{\cal L} &=& - 8c_1a_1 \epfour \bar{\Psi}_{\mu\nu}
\Gamma^{abce} \Psi_{\alpha\beta} \hat{\eta}^{de} \\
&& - i(12c_2a_1-3c_4) \epthree \bar{\Psi}_{\mu\nu} \Gamma^{abd}
\Phi_\alpha \hat{\eta}^{cd} + (8c_3a_1-2c_5) \eptwo \bar{\Phi}_\mu
\Gamma^{ac} \Phi_\nu \hat{\eta}^{bc}
\end{eqnarray*}
The first term vanishes in $d=4$ due to the identity
$$
0 = \epfive \bar{\Psi}_{\mu\nu} \Gamma^{abcd} {\Psi}_{\alpha\beta}
\hat{\eta}_\gamma{}^{e} = 4 \epfour \bar{\Psi}_{\mu\nu}
\Gamma^{abce} {\Psi}_{\alpha\beta} \hat{\eta}^{de}
$$
while the last two terms give restrictions
\begin{equation}\label{CondGrIntParamMassv2}
c_4 = 4c_2a_1, \qquad c_5 = 4c_3a_1
\end{equation}

At last we consider variations under the
$\hat{\xi}^a$-transformations:
\begin{eqnarray*}
\delta{\cal L} &=& -i(16c_1a_1M+\frac32mc_4) \epfour
\bar{\Psi}_{\mu\nu} \Gamma^{abc} \Psi_{\alpha\beta} \hat{\xi}^d  \\
&& +(32c_1a_1m+3mc_5) \epthree \bar{\Psi}_{\mu\nu} \Gamma^{ab}
\Phi_{\alpha} \hat{\xi}^c  \\
&& + (-4Ma_1c_2+Mc_4) \epthree \bar{\Psi}_{\mu\nu}\Gamma^{abcd}
\Phi_\alpha \hat{\xi}^d  \\
&& -i(4a_1c_2m+4c_3Ma_1-mc_4-Mc_5) \eptwo \bar{\Phi}_{\mu}
\Gamma^{abc} \Phi_\nu \hat{\xi}^c  \\
&& + i(16a_1c_2m-8Mc_3a_1+2mc_4-4Mc_5) \eptwo \bar{\Phi}_{\mu}
\gamma^a \Phi_\nu \hat{\xi}^b
\end{eqnarray*}
Taking into account (\ref{CondLagParam}), (\ref{CondGrIntParam}),
(\ref{CondGrIntParamMassv1}), (\ref{CondGrIntParamMassv2}) we obtain
simply:
$$
\delta {\cal L} = - \frac{i}{2} \epfour
\bar{\Psi}_{\mu\nu}\Gamma^{abc} \Psi_{\alpha\beta} \hat{\xi}^d + 2ia_1
\eptwo \bar\Phi_{\mu} \gamma^a \Phi_\nu \hat{\xi}^b
$$
To compensate these variations one can use the following corrections
$$
\delta_1 \psi_\mu \sim {\Psi}_{\mu\nu} \hat{\xi}^\nu, \qquad \delta_1
\phi \sim \Phi_\mu \hat{\xi}^\mu
$$

Thus the requirement that the Lagrangian be gauge invariant fixes
the coefficients $c_{4,5}$ for the Abelian vertices and also relates
coefficients $b_1$ and $c_1$ (which is just a manifestation of
universality of gravitational interaction). Note that by Metsaev
classification \cite{Met07b} in general dimensions $d > 4$ we would
have two non-minimal vertices with two derivatives as well as the
minimal one. But in $d=4$ dimensions these two derivative vertices
are absent (on-shell and up to possible field redefinitions) as we
will now show.

\noindent
{\bf Vertex $\frac32-\frac32-2$.} For this vertex we
obtain:
$$
{\cal L} = (-48c_1a_1+8c_0b_1) \epfour D_\mu \omega_{\nu}{}^{ab}
\bar{\psi}_{\alpha} \Gamma^{cd} \psi_{\beta} = 0
$$
This expression vanishes due to obtained expressions and
restrictions for the arbitrary coefficients

\noindent {\bf Vertex $\frac12-\frac12-2$.} In this case we have
\begin{eqnarray*}
{\cal L} &=& (-c_3a_1+8c_0b_5) \eptwo D_\mu \omega_\nu{}^{cd}
\bar{\phi} \Gamma^{abcd} \phi + (8c_3a_1-2c_5) \eptwo
\omega_\mu{}^{ac} D_\nu \bar{\phi} \Gamma^{bc} \phi  \\
&& + (2c_3a_1+16c_0b_4) \eptwo D_\mu \omega_\nu{}^{ab} \bar{\phi} \phi
\end{eqnarray*}
The second term drops out due to relations among the coefficients,
the first term vanishes on-shell for gravitational field due to
identity
$$
0 = \epthree R_{\mu\nu,\alpha}{}^d \bar{\phi} \Gamma^{abcd} \phi =
2 \epthree (D_\mu \omega_{\nu,\alpha}{}^d + \lambda^2
e_\mu{}^d h_{\nu,\alpha}) \bar{\phi} \Gamma^{abcd} \phi =
6 \eptwo D_\mu \omega_{\nu}{}^{cd} \bar{\phi} \Gamma^{abcd} \phi
$$
while the third term can be removed by field redefinition
$$
h_\mu{}^a \rightarrow h_\mu{}^a + \kappa_3 e_\mu{}^a \bar{\phi} \phi
$$
\noindent
{\bf Vertex $\frac32-\frac12-2$.} For the last possibility
we get:
\begin{eqnarray*}
{\cal L} &=& i(-24c_2a_1+6c_4) \epthree \omega_\alpha{}^{da} D_{\mu}
\bar{\psi}_{\nu} \Gamma^{bcd} \phi + \\
&& +i(16c_0b_2-12c_2a_1) \epthree D_\nu \omega_\alpha{}^{ab}
\bar{\psi}_\mu \gamma^c \phi - i(16c_0b_3-12c_2a_1) \epthree
D_\nu \omega_\alpha{}^{dc} \bar{\psi}_\mu \Gamma^{abd} \phi
\end{eqnarray*}
The first term drops out due to relations among the coefficients,
the last term vanishes on-shell for gravitational field, and the
second term can be removed by the field redefinition
$$
h_\mu{}^a \rightarrow h_\mu{}^a + \kappa_2 \bar{\psi}_\mu \gamma^a
\phi
$$

As a result, the expression (\ref{gvertex}) is the final form for
Lagrangian of coupled spin-2 and spin-3/2 fields in $AdS_4$ space
including the cubic interaction vertex. The parameters $c_1, c_2,
c_3$ are fixed in the free gravitational field Lagrangian
(\ref{FreeLagS2}), the parameter $c_0$ is given by
(\ref{CurvLagS2}). The parameters $c_4, c_5$ and the parameters of
gauge transformations are expressed through the single parameter
$a_1$ which is the only free parameter of the theory.

\section*{Conclusion}

In this paper we have constructed the cubic vertices for massive
spin-3/2 field coupled to electromagnetic and gravitational fields
in $AdS_4$ space. The corresponding vertices are gauge invariant due
to presence of Stueckelberg auxiliary fields and contain some number
of free parameters. The results are given by the expressions
(\ref{emvertex}) or (\ref{emavertex}) for electromagnetic
interaction and (\ref{gvertex}) for gravitational interaction.

Construction of vertices is based on generalization of
Fradkin-Vasiliev formalism \cite{FV87,FV87a} where the main building
blocks of Lagrangians are the gauge invariant curvatures. Although
this formalism was known only for massless higher spin theories, we
have shown, on the example of spin-3/2 field, that the gauge
invariant curvatures can in principle be constructed for massive
higher spin fields as well. Certainly the case of massive fermionic
fields appears to be the most technically involved one. Nevertheless
it seem worth to apply this formalism to massive fermionic fields
with spins higher than 3/2.

As we pointed out, the constructed vertices contain some number of
arbitrary parameters (two for electromagnetic coupling and one for
gravitational coupling) which can not be fixed only form gauge
invariance. In principle one can hope that the additional
constraints for those parameters can appear from requirements of
causality. The possibility of such constraints was already
demonstrated in the papers \cite{BSZ12,BDT13}.

\section*{Acknowledgments}
I.L.B and T.V.S are grateful to the grant for LRSS, project No.
88.2014.2 and RFBR grant, project No. 12-02-00121-a for partial
support. Their research was also supported by grant of Russian
Ministry of Education and Science, project TSPU-122. TVS
acknowledges partial support from RFBR grant No. 14-02-31254. Work
of Yu.M.Z was supported in parts by RFBR grant No. 14-02-01172.


\begin{thebibliography}{10}

\bibitem{Vas01}
M.~A. Vasiliev
{\it "Cubic Interactions of Bosonic Higher Spin Gauge Fields in
$AdS_5$",}
Nucl.Phys. {\bf B616} (2001) 106-162; Erratum-ibid. B652 (2003) 407,
arXiv:hep-th/0106200.

\bibitem{AV02}
K.~B. Alkalaev, M.~A. Vasiliev
{\it "N=1 Supersymmetric Theory of Higher Spin Gauge Fields in AdS(5)
at the Cubic Level",}
Nucl.Phys. {\bf B655} (2003) 57-92, arXiv:hep-th/0206068.

\bibitem{BFPT06}
I.~L. Buchbinder, A. Fotopoulos, A.~C. Petkou, M. Tsulaia
{\it
"Constructing the cubic interaction vertex for higher spin gauge
fields",}
Phys. Rev. {\bf D74} (2006) 105018, arXiv:hep-th/0609082.

\bibitem{BL06}
N.~Boulanger, S.~Leclercq
{\it "Consistent couplings between spin-2 and spin-3 massless
fields",}
JHEP {\bf 0611} (2006) 034, arXiv:hep-th/0609221.

\bibitem{BLS08}
N.~Boulanger, S.~Leclercq, P.~Sundell
{\it "On The Uniqueness of Minimal Coupling in Higher-Spin Gauge
Theory",}
JHEP {\bf 0808} (2008) 056, arXiv:0805.2764.

\bibitem{Zin10}
Yu.~M. Zinoviev
{\it "Spin 3 cubic vertices in a frame-like formalism",}
JHEP {\bf 08} (2010) 084, arXiv:1007.0158.

\bibitem{MMR10a}
R.~Manvelyan, K.~Mkrtchyan, W.~Ruehl
{\it "General trilinear interaction for arbitrary even higher spin
gauge fields",}
Nucl. Phys. {\bf B836} (2010) 204, arXiv:1003.2877.

\bibitem{MMR10b}
R.~Manvelyan, K.~Mkrtchyan, W.~Ruehl
{\it "A generating function for the cubic interactions of higher spin
fields",}
Phys. Lett. {\bf B696} (2011) 410, arXiv:1009.1054.

\bibitem{JT11}
E.~Joung, M.~Taronna
{\it "Cubic interactions of massless higher spins in (A)dS:
metric-like approach",}
Nucl. Phys. {\bf B861} (2012) 145, arXiv:1110.5918.

\bibitem{Met05}
R.~R. Metsaev
{\it "Cubic interaction vertices of massive and massless higher spin
fields",}
Nucl. Phys. {\bf B759} (2006) 147, arXiv:hep-th/0512342.

\bibitem{Met07b}
R.~R. Metsaev
{\it "Cubic interaction vertices for fermionic and bosonic arbitrary
spin fields",}
Nucl. Phys. {\bf B859} (2012) 13, arXiv:0712.3526.

\bibitem{Met12}
R.~R. Metsaev
{\it "BRST-BV approach to cubic interaction vertices for massive and
massless higher-spin fields",}
Phys. Lett. {\bf B720} (2013) 237, arXiv:1205.3131.

\bibitem{Zin06}
Yu.~M. Zinoviev
{\it "On massive spin 2 interactions",}
Nucl. Phys. {\bf B770} (2007) 83-106, arXiv:hep-th/0609170.

\bibitem{Zin08}
Yu.~M. Zinoviev
{\it "On spin 3 interacting with gravity",}
Class. Quantum Grav. {\bf 26} (2009) 035022, arXiv:0805.2226.

\bibitem{Zin09}
Yu.~M. Zinoviev
{\it "On massive spin 2 electromagnetic interactions",}
Nucl. Phys. {\bf B821} (2009) 431-451, arXiv:0901.3462.

\bibitem{ST10}
A.~Sagnotti, M.~Taronna
{\it "String Lessons for Higher-Spin Interactions",}
Nucl. Phys. {\bf B842} (2011) 299, arXiv:1006.5242.

\bibitem{BSZ12}
I.~L. Buchbinder, T.~V. Snegirev, Yu.~M. Zinoviev
{\it "Cubic interaction vertex of higher-spin fields with external
  electromagnetic field",}
Nucl. Phys. {\bf B864} (2012) 694, arXiv:1204.2341.

\bibitem{Zin12}
Yu.~M. Zinoviev
{\it "On massive gravity and bigravity in three dimensions",}
Class. Quant. Grav. {\bf 30} (2013) 055005, arXiv:1205.6892.

\bibitem{BSZ12b}
I.~L. Buchbinder, T.~V. Snegirev, Yu.~M. Zinoviev
{\it "On gravitational interactions for massive higher spins in
$AdS_3$",}
J. Phys. A {\bf 46} (2013) 214015, arXiv:1208.0183.

\bibitem{BDT13}
I.~L. Buchbinder, P.~Dempster, M.~Tsulaia {\it "Massive Higher Spin
Fields Coupled to a Scalar: Aspects of Interaction and Causality",}
Nucl. Phys. {\bf B877} (2013) 260, arXiv:1308.5539.

\bibitem{Zin13}
Yu.~M. Zinoviev {\it "Massive spin-2 in the Fradkin-Vasiliev
formalism. I. Partially massless case",} arXiv:1405.4065.

\bibitem{Met06}
R.~R. Metsaev
{\it "Gravitational and higher-derivative interactions of massive spin
5/2 field in (A)dS space",}
Phys. Rev. {\bf D77} (2008) 025032, arXiv:hep-th/0612279.

\bibitem{GHR12}
G.~L. Gomez, M.~Henneaux, R.~Rahman
{\it "Higher-Spin Fermionic Gauge Fields and Their Electromagnetic
Coupling",}
JHEP {\bf 1208} (2012) 093, arXiv:1206.1048.

\bibitem{HGR13}
M. Henneaux, G.L. Gomez, R. Rahman {\it "Gravitational Interactions
of Higher-Spin Fermions",} JHEP {\bf 1401} (2014) 087,
arXiv:1310.5152.

\bibitem{FV87}
E.~S. Fradkin, M.~A. Vasiliev
{\it "On the gravitational interaction of massless higher-spin
fields",}
Phys. Lett. {\bf B189} (1987) 89.

\bibitem{FV87a}
E.~S. Fradkin, M.~A. Vasiliev
{\it "Cubic interaction in extended theories of massless higher-spin
fields",}
Nucl. Phys. {\bf B291} (1987) 141.

\bibitem{Alk10}
K.B. Alkalaev
{\it "FV-type action for AdS(5) mixed-symmetry fields",}
JHEP {\bf 1103} (2011) 031, arXiv:1011.6109.

\bibitem{Zin10a}
Yu.~M. Zinoviev
{\it "On electromagnetic interactions for massive mixed symmetry
field",}
JHEP {\bf 03} (2011) 082, arXiv:1012.2706.

\bibitem{BSZ11}
N. Boulanger, E.~D. Skvortsov, Yu.~M. Zinoviev {\it "Gravitational
cubic interactions for a simple mixed-symmetry gauge field in AdS
and flat backgrounds",} J. Phys. {\bf A44} (2011) 415403,
arXiv:1107.1872.

\bibitem{Zin11}
Yu.~M. Zinoviev
{\it "Gravitational cubic interactions for a massive mixed symmetry
gauge field",}
Class. Quantum Grav. {\bf 29} (2012) 015013, arXiv:1107.3222.

\bibitem{BS11}
N. Boulanger, E.~D. Skvortsov {\it "Higher-spin algebras and cubic
interactions for simple mixed-symmetry fields in AdS spacetime",}
JHEP {\bf 1109} (2011) 063, arXiv:1107.5028.

\bibitem{BPS12}
N. Boulanger, D. Ponomarev, E.D. Skvortsov {\it "Non-abelian cubic
vertices for higher-spin fields in anti-de Sitter space",} JHEP {\bf
1305} (2013) 008, arXiv:1211.6979.

\bibitem{Vas11}
M.~Vasiliev
{\it "Cubic Vertices for Symmetric Higher-Spin Gauge Fields in
$(A)dS_d$",}
Nucl. Phys. {\bf B862} (2012) 341, arXiv:1108.5921.

\bibitem{DPW00}
S. Deser, V. Pascalutsa, A. Waldron
{\it "Massive Spin 3/2 Electrodynamics",}
Phys.Rev. {\bf D62} (2000) 105031, arXiv:hep-th/0003011.

\bibitem{DW01d}
S. Deser, A, Waldron
{\it "Inconsistencies of Massive Charged Gravitating Higher Spins",}
Nucl.Phys. {\bf B631} (2002) 369, arXiv:hepth/0112182.

\bibitem{Rah11}
R.~Rahman
{\it "Helicity-1/2 Mode as a Probe of Interactions of Massive
Rarita-Schwinger Field",}
Phys. Rev. {\bf D87} (2013) 065030, arXiv:1111.3366.

\end{thebibliography}
\end{document}